\documentclass[twocolumn,aps,prl] {revtex4} 
\usepackage{graphicx}
\begin{document}
\pagestyle{empty} 
\title{On the dependence of the leak-rate of seals on the skewness of the surface height probability distribution}
\author{  
B. Lorenz$^{1,2}$ and B.N.J. Persson$^1$} 
\affiliation{$^1$ IFF, FZ J\"ulich, D-52425 J\"ulich, Germany}
\affiliation{$^2$ IFAS, RWTH Aachen University, D-52074 Aachen, Germany}

\begin{abstract}
Seals are extremely useful devices to prevent fluid leakage. 
We present experimental result which show that the leak-rate of seals depend
sensitively on the skewness in the height probability distribution. The experimental data are analyzed
using the critical-junction theory.
We show that using the top-power spectrum result in good agreement between theory and experiment.
\end{abstract}
\maketitle


A seal is a device for closing a gap or making a joint fluid tight\cite{Flitney}.
Seals play a crucial role in many modern engineering devices, and the failure of
seals may result in catastrophic events, such as the Challenger disaster. 
In spite of its apparent 
simplicity, it is not easy to predict the leak-rate and
(for dynamic seals) the friction forces\cite{Mofidi}. 
The main problem is the influence of surface
roughness on the contact mechanics at the seal-substrate interface. 
Most surfaces of engineering interest have surface roughness
on a wide range of length scales\cite{P3}, e.g, from cm to nm, which will influence the leak rate
and friction of seals, and accounting for the whole range of surface roughness
is impossible using standard numerical methods, such as the Finite Element Method.

\begin{figure}
\includegraphics[width=0.35\textwidth,angle=0.0]{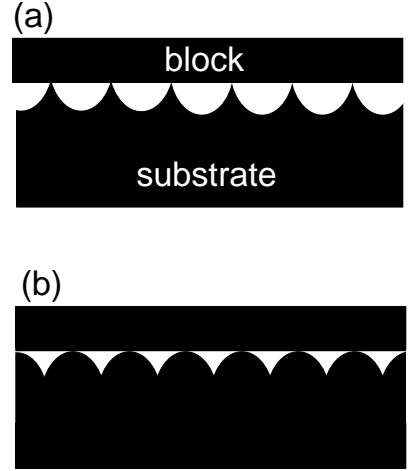}
\caption{\label{twotypes}
Contact between a rigid block with a flat surface and a rigid substrate
with periodic surface structures.
The two substrate surfaces in (a) and (b) have the same surface roughness power spectrum.
Note that the empty volume between the surfaces is much larger in the case (a) than
in case (b).}
\end{figure}

Randomly rough surfaces have Gaussian height probability distribution but many surfaces of engineering
interest have skewed distributions which may effect the leak rate of seals. 
To illustrate this we consider an extreme case: a rigid solid block with a flat surface in contact
with a rigid substrate with periodic ``roughness'' as in Fig. \ref{twotypes}. 
The substrate surfaces in (a) and (b) have the same root-mean-square roughness and the same 
surface roughness power spectrum, but it is clear that in (a) the empty volume between the surfaces is larger than in (b), resulting 
in a larger leak rate. In the real situation the roughness is not periodic and the solids are
not rigid, but one may expect a higher leak rate for the situation
where the asymmetry of the height profile is as for case (a). 

To study the point discussed above, we have performed experiments using sandpaper which has a skewed
height probability distribution as in Fig. \ref{twotypes}(a). We have also used 
surfaces with ``inverted'' surface roughness profile by producing a ``negative'' of the sandpaper surface 
using silicon rubber. In the latter experiments we squeezed a silicon rubber ring, which was 
cross-linked with the sandpaper surface as the substrate, against a flat glass surface. 
By comparing the measured leak-rate for this configuration with that for a silicon ring with
flat bottom surface squeezed against the same sandpaper surface, we are able to address the
problem illustrated in Fig. \ref{twotypes}.

\begin{figure}
\includegraphics[width=0.22\textwidth,angle=0.0]{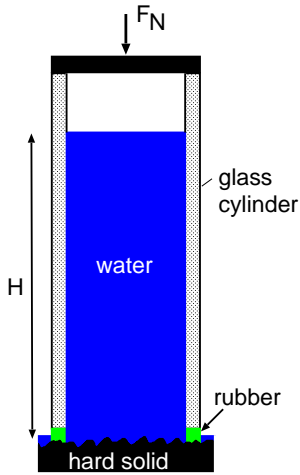}
\caption{\label{tube.water}
Experimental set-up for measuring the leak-rate of seals.
A glass (or PMMA) cylinder with a rubber ring attached to one end is squeezed against
a hard substrate with well-defined surface roughness. The cylinder is filled with 
water, and the leak-rate of the water at the rubber-countersurface
is detected by the change in the height of the water in the cylinder. 
}
\end{figure}

We briefly describe the leak-rate model\cite{Creton,LorenzEPL,P3,Yang,LP,Carbone} and experimental set-up used in this study.
Consider the fluid leakage through a rubber seal, from a high fluid pressure $P_{\rm a}$ region, to a
low fluid pressure $P_{\rm b}$ region. 
In our experimental study we have used 
the experimental set-up shown in Fig.  \ref{tube.water} for measuring the leak-rate of seals.
A glass (or PMMA) cylinder with a rubber ring attached to one end is squeezed against
a hard substrate with well-defined surface roughness. The cylinder is filled with 
water, and the leak-rate of the water at the rubber-countersurface
is detected by the change in the height of the water in the cylinder. 
Thus $P_{\rm a}-P_{\rm b}=\rho g H$, where $H$ is the height of the water column, and
$\rho$ the mass density of water. For further experimental details, see Ref. \cite{LorenzEPL,LP}.

Assume that the nominal contact region between the rubber and the hard countersurface is
rectangular with area $L_x\times L_y$, with $L_y > L_x$. 
We assume that the high pressure fluid region is for $x<0$
and the low pressure region for $x>L_x$. We  ``divide'' the contact region into squares with
the side $L_x=L$ and the area $A_0=L^2$ (this assumes that $N=L_y/L_x$ is an integer, but this 
restriction does not affect the final result). 
Now, let us study the contact between the two solids within one of the squares
as we increase the magnification $\zeta$. We define $\zeta= L/\lambda$, where $\lambda$ is the resolution.
We study how the apparent contact area (projected on the $xy$-plane),
$A(\zeta)$, between the two solids depends on the magnification $\zeta$.
At the lowest magnification we cannot observe any surface roughness, and 
the contact between the solids appears to be complete i.e., $A(1)=A_0$. 
As we increase the magnification
we will observe some interfacial roughness, and the (apparent) contact area will decrease.
At high enough magnification, say $\zeta = \zeta_{\rm c}$, a percolating path of 
non-contact area will be observed 
for the first time. 
We denote the most narrow constriction along this percolation path as
the {\it critical constriction}. The critical constriction will have the lateral
size $\lambda_{\rm c} = L/\zeta_{\rm c}$ and the surface separation at this point is denoted by 
$u_{\rm c}$. We can calculate $u_{\rm c}$
using a recently developed contact mechanics theory\cite{YangPersson} (see below). 
As we continue to increase the magnification we will find more percolating channels 
between the surfaces, but these will have more narrow constrictions 
than the first channel which appears at $\zeta=\zeta_{\rm c}$, and as a first approximation one may
neglect the contribution to the leak-rate from these channels\cite{Yang}. 

A first rough estimate of the leak-rate is obtained by assuming that all the leakage 
occurs through the critical percolation channel, and that
the whole pressure drop $\Delta P = P_{\rm a}-P_{\rm b}$ (where $P_{\rm a}$ and $P_{\rm b}$ is the 
pressure to the left and right of the
seal) occurs over the critical constriction (of width and length $\lambda_{\rm c} \approx L/\zeta_{\rm c}$
and height $u_{\rm c}$). 
We refer to this theory as the {\it critical-junction} theory.
If we approximate the critical constriction
as a pore with rectangular cross section (width and length $\lambda_c$ and height $u_c << \lambda_c$),  
and if we assume an incompressible
Newtonian fluid, the volume-flow per unit time through the critical constriction
will be given by (Poiseuille flow) 
$$\dot Q = {u_c^3 \over 12 \eta}  \Delta P,\eqno(1)$$
where $\eta $ is the fluid viscosity. 
In deriving (1) we have assumed laminar flow and that $u_c << \lambda_c$,
which is always satisfied in practice. 
Finally, since there are
$N=L_y/L_x$ square areas in the rubber-countersurface (apparent) contact area, we get the total leak-rate
$$\dot Q = {L_y \over L_x} {u_c^3 \over 12 \eta}  \Delta P.\eqno(2)$$ 
Note that a given percolation channel could have several narrow (critical or nearly critical)
constrictions of nearly the same dimension
which would reduce the flow along the channel. But in this case one would also expect more channels from
the high to the low fluid pressure side of the junction, which would tend to increase the leak rate.
These two effects will, at least in the simplest picture, compensate each other (see Ref. \cite{Yang}). 
The effective medium theory presented in Ref. \cite{LP} 
includes (in an approximate way) all the flow channels, but gives results very similar to the critical-junction theory
described above\cite{cpath}.

To complete the theory we must calculate the separation $u_{\rm c}$ 
of the surfaces at the
critical constriction. We first determine the critical magnification $\zeta_{\rm c}$ by assuming that the 
apparent relative contact area at this point is given by percolation theory. 
Thus, the relative contact area $A(\zeta)/A_0 \approx 1-p_{\rm c}$, where $p_{\rm c}$  is the 
so called percolation threshold\cite{Stauffer}. 
Numerical contact mechanics studies, such as those presented in Ref. \cite{Yang} and Ref. \cite{thesis},
typically give $p_{\rm c}$ between $0.5$ and $0.6$. 
For finite sized systems the percolation will, on the average, occur for (slightly) smaller values
of $p_{\rm c}$, and fluctuations in the percolation threshold will occur between 
different realizations of the same physical system. 
Here we use $p_{\rm c} = 0.6$ to determine the critical magnification $\zeta=\zeta_{\rm c}$.

\begin{figure}
\includegraphics[width=0.35\textwidth,angle=0.0]{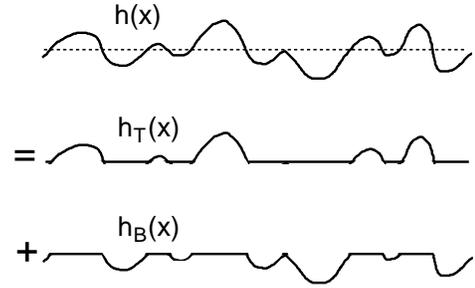}
\caption{\label{bottomtopxx}
The surface profile $h(x)$ is decomposed into a top $h_{\rm T}(x)$ and a bottom
$h_{\rm B}(x)$ profile.
}
\end{figure}

\begin{figure}
\includegraphics[width=0.45\textwidth,angle=0.0]{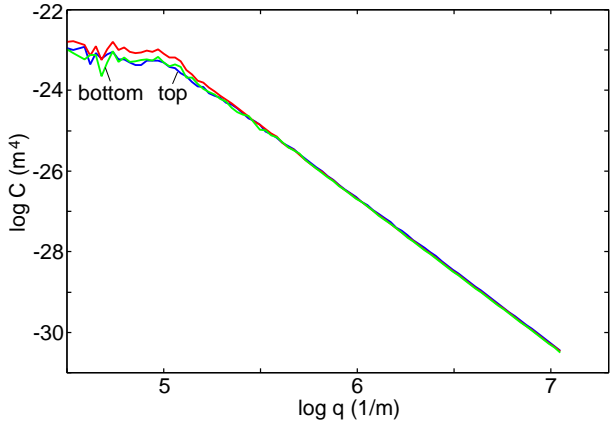}
\caption{\label{Mathematical.Full.T.B}
The power spectrum $C(q)$ and the top $C^*_{\rm T}(q)$ and bottom $C^*_{\rm B}(q)$ power spectrum
for a mathematically generated randomly rough surface with a Gaussian height probability
distribution. The surface is self affine fractal for $q>10^5 \ {\rm m}^{-1}$ with the
fractal dimension $D_{\rm f}=2.2$ and the root-mean-square roughness $0.8 \ {\rm \mu m}$.
}
\end{figure}

The (apparent) relative contact area $A(\zeta)/A_0$ and the interfacial separation $u_1(\zeta)$ 
at the magnification $\zeta$ can be obtained using the contact mechanics 
formalism developed elsewhere\cite{PSSR,PerssonPRL,YangPersson,P1,Bucher,JCPpers,Lorenz,PerssonJPCM}.
We define $u_1(\zeta)$ to be the (average) height separating the surfaces which appear to come into 
contact when the magnification decreases from $\zeta$ to $\zeta-\Delta \zeta$, where $\Delta \zeta$
is a small (infinitesimal) change in the magnification. 
Since the surfaces of the solids are everywhere rough the actual 
separation between the solid walls 
will fluctuate around the average $u_1(\zeta)$. Thus we expect $u_c=\alpha u_1(\zeta_c)$, where
$\alpha < 1$ (but of order unity). 
We note that $\alpha$ is due to the surface 
roughness which occur at length scales shorter
than $\lambda_c$, see Ref. \cite{LP}.

In the contact mechanics theory of Persson, the surface roughness enter only via the surface roughness power spectrum
$$C(q) = {1\over (2\pi)^2} \int d^2x \ \langle h({\bf x})h({\bf 0})\rangle {\rm e}^{-i{\bf q}\cdot {\bf x}}$$
where $\langle ... \rangle$ stands for ensemble average, and where we have assumed 
$\langle h \rangle = 0$. A randomly rough surface has a Gaussian 
height probability distribution, $P(h)$, but many surfaces of practical use have a skewed height distribution.
For this latter case it is useful to introduce the {\it top} and {\it bottom} power spectra defined as follows\cite{P3}: 
$$C_{\rm T}(q) = {1\over (2\pi)^2} \int d^2x \ \langle h_{\rm T}({\bf x})h_{\rm T}({\bf 0})\rangle {\rm e}^{-i{\bf q}\cdot {\bf x}}$$
$$C_{\rm B}(q) = {1\over (2\pi)^2} \int d^2x \ \langle h_{\rm B}({\bf x})h_{\rm B}({\bf 0})\rangle {\rm e}^{-i{\bf q}\cdot {\bf x}}$$
where $h_{\rm T}({\bf x})= h({\bf x})$ for $h > 0$ and zero otherwise, while $h_{\rm B}({\bf x})= h({\bf x})$ for $h < 0$ and zero otherwise.
These are ``rectified'' profiles; see Fig. \ref{bottomtopxx}. It is clear by symmetry that for a randomly rough surface with
Gaussian height distribution, $C_{\rm T}(q)=C_{\rm B}(q)$. If $n_{\rm T}$ and $n_{\rm B}$ are the fractions of the nominal surface area (i.e., the surface area projected
on the $xy$-plane) where $h>0$ and $h<0$, respectively, then we also define 
$C^*_{\rm T}(q) = C_{\rm T}(q)/n_{\rm T}$ and $C^*_{\rm B}(q) = C_{\rm B}(q)/n_{\rm B}$. Roughly speaking, $C^*_{\rm T}$ would be the power spectrum resulting if the
actual bottom profile (for $h<0$) was replaced by a mirrored top profile (for $h>0$). A similar statement holds for $C^*_{\rm B}$. 
For randomly rough surfaces with Gaussian height distribution we expect $C^*_{\rm T}(q)=C^*_{\rm B}(q) \approx C(q)$. That this is indeed the case
is illustrated in Fig. \ref{Mathematical.Full.T.B} which shows the calculated $C(q)$, $C^*_{\rm T}(q)$ and $C^*_{\rm B}(q)$ 
for a mathematically generated randomly rough surface with a Gaussian height probability
distribution. The surface is self affine fractal for $q> 10^5 \ {\rm m}^{-1}$ with the
fractal dimension $D_{\rm f}=2.2$ and the root-mean-square roughness $0.8 \ {\rm \mu m}$.

The contact mechanics theory of Persson can be applied approximately to surfaces with skewed 
height distribution. However in this case, at least for small squeezing pressures where the contact only 
occur at the highest asperities, one should use $C^*_{\rm T}(q)$ rather than $C(q)$ in order to better represent the surface roughness. We will show below
that by using $C^*_{\rm T}(q)$ we can quantitatively understand the leak-rate of rubber seals squeezed against surfaces with skewed
height probability distribution.   

\begin{figure}
\includegraphics[width=0.45\textwidth,angle=0.0]{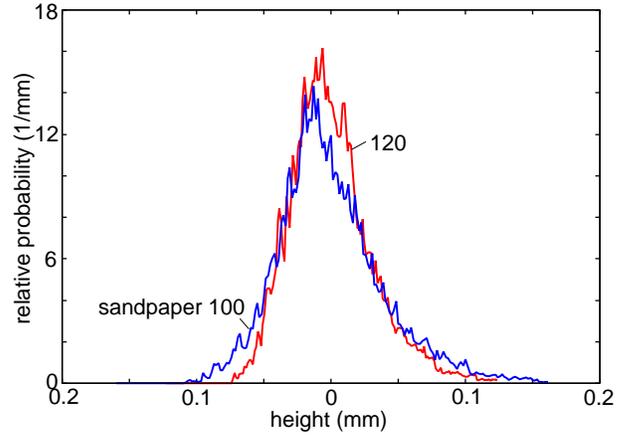}
\caption{\label{Ph.sandpaper100.120}
The surface height probability distribution for sandpaper 100 and 120
surfaces with the root-mean-square roughness amplitudes $40 \ {\rm \mu m}$ and
$31 \ {\rm \mu m}$. The two surfaces have the skewness $\langle h^3 \rangle /\langle h^2 \rangle^{3/2} = 0.85$ and $0.82$
respectively.
}
\end{figure}

\begin{figure}
\includegraphics[width=0.45\textwidth,angle=0.0]{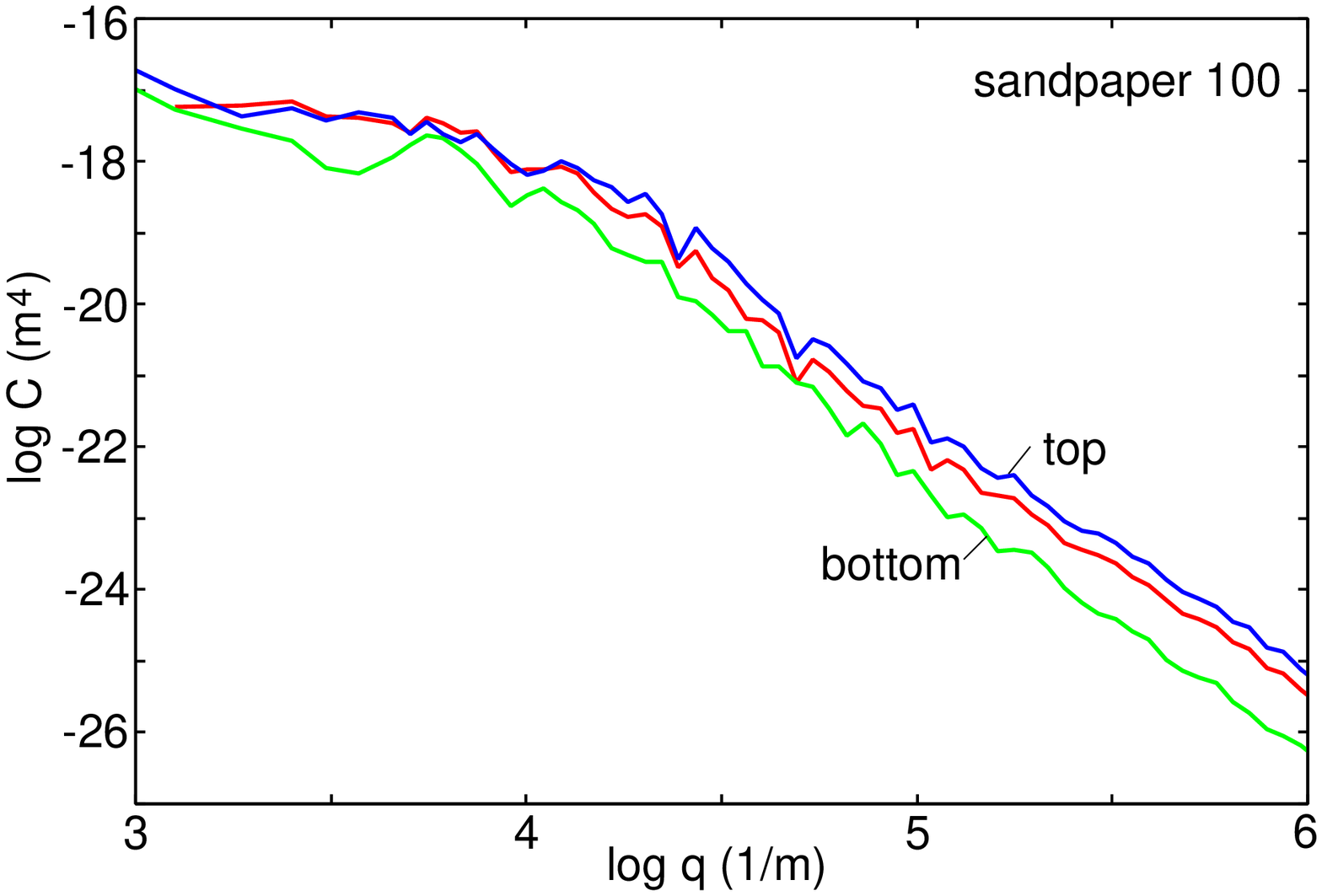}
\caption{\label{Cq.sandpaper100}
Surface roughness power spectrum of 
sandpaper 100 surface. 
The three curves are the surface roughness power spectrum $C(q)$
of the original surface (red line), and the top $C^*_{\rm T}(q)$ (blue) 
and bottom $C^*_{\rm B}(q)$ (green) surface roughness power spectrum.
The surface has the root-mean-square roughness
$40 \ {\rm \mu m}$. 
The fraction of the (projected) surface area above above the average plane is about $0.44$.
}
\end{figure}

\begin{figure}
\includegraphics[width=0.45\textwidth,angle=0.0]{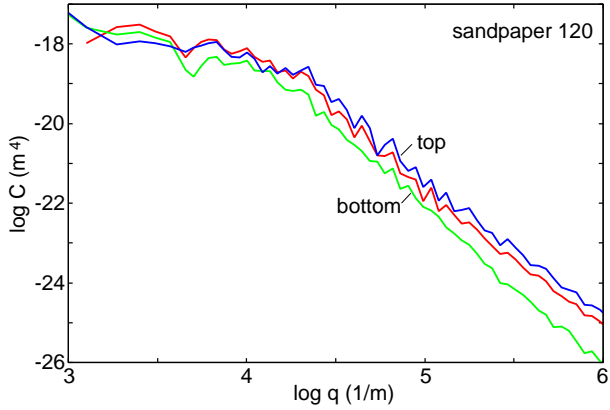}
\caption{\label{Cq.sandpaper120}
Surface roughness power spectrum of 
sandpaper 120 surface. 
The three curves are the surface roughness power spectrum $C(q)$
of the original surface (red line), and the top $C^*_{\rm T}(q)$ (blue) 
and bottom $C^*_{\rm B}(q)$ (green) surface roughness power spectrum.
The surface has the root-mean-square roughness
$31 \ {\rm \mu m}$. 
The fraction of the (projected) surface area above above the average plane is about $0.45$.
}
\end{figure}

We have performed experiments using 
two sandpaper surfaces (corundum paper, grit size 100 and 120) 
with the the root-mean-square roughness $40 \ {\rm \mu m}$ and $31 \ {\rm \mu m}$.
From the measured surface topography we obtain the height probability distribution $P(h)$ 
(Fig. \ref{Ph.sandpaper100.120}) and the surface roughness power spectrum 
shown in Fig. \ref{Cq.sandpaper100} and \ref{Cq.sandpaper120}, respectively.  
Note that for both surfaces $P(h)$ is asymmetric with a tail towards higher $h$.
This is easy to understand: sandpaper surfaces consist of
particles with sharp edges pointing above the surface, while the region between 
the particles is filled with a resin-binder making the valleys 
smoother and wider than the peaks [as in Fig. \ref{twotypes}(a)], which result
in an asymmetric $P(h)$ as observed. 

\begin{figure}
\includegraphics[width=0.45\textwidth,angle=0.0]{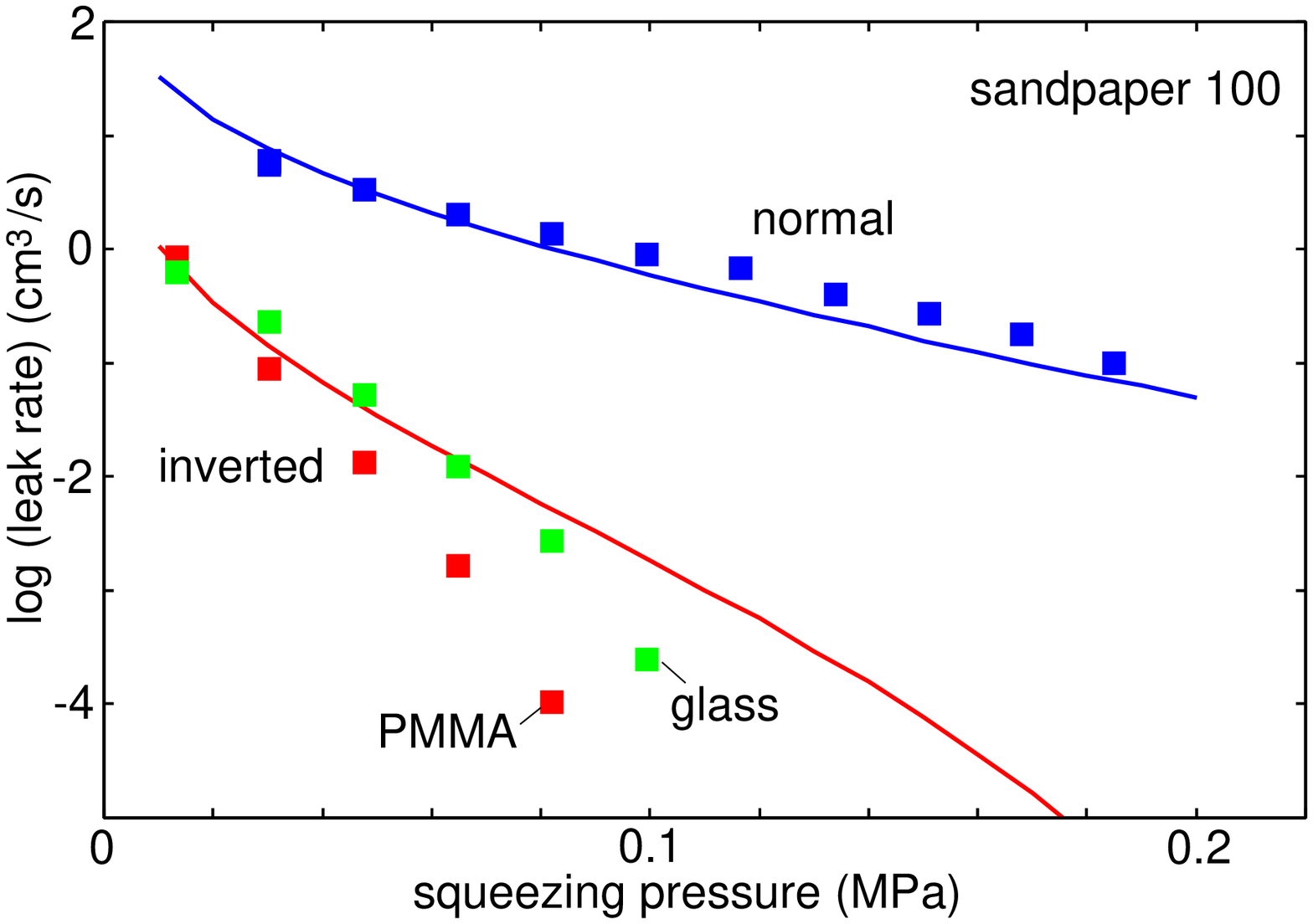}
\caption{\label{Sandpaper100.normal.alpha=1.inverted.alpha=0.8}
Square symbols: the measured leak rate for sandpaper 100 substrate (upper symbols) and for
an inverted surface (lower symbols). 
The solid lines are
the calculated leak rate using the critical-junction theory with the percolation 
threshold $p_{\rm c} = 0.6$. In the
calculation for the top curve we used the top power spectrum $C^*_{\rm T}(q)$ obtained from the
measured surface topography. For the inverted surface (bottom curve) we used the bottom
power spectrum $C^*_{\rm B}(q)$. The measured rubber elastic
modulus $E=2.3 \ {\rm MPa}$ and the fluid pressure difference 
$\Delta P = P_{\rm a}-P_{\rm b} = 10 \ {\rm kPa}$ obtained from the height of the water column.  
In the calculations we have used $\alpha = 1$ (upper curve) and $\alpha = 0.8$ (lower curve).
}
\end{figure}

In Fig. \ref{Sandpaper100.normal.alpha=1.inverted.alpha=0.8}
we show the measured leak rate for sandpaper 100 substrate (upper squares) and for
an inverted surface (lower squares). 
The solid lines are
the calculated leak rate using the critical-junction theory. In the
calculation for the top curve we used the top power spectrum $C^*_{\rm T}(q)$ obtained from the
measured surface topography. For the inverted surface (bottom curve) we used the bottom
power spectrum $C^*_{\rm B}(q)$ of the sandpaper surface. 

\begin{figure}
\includegraphics[width=0.45\textwidth,angle=0.0]{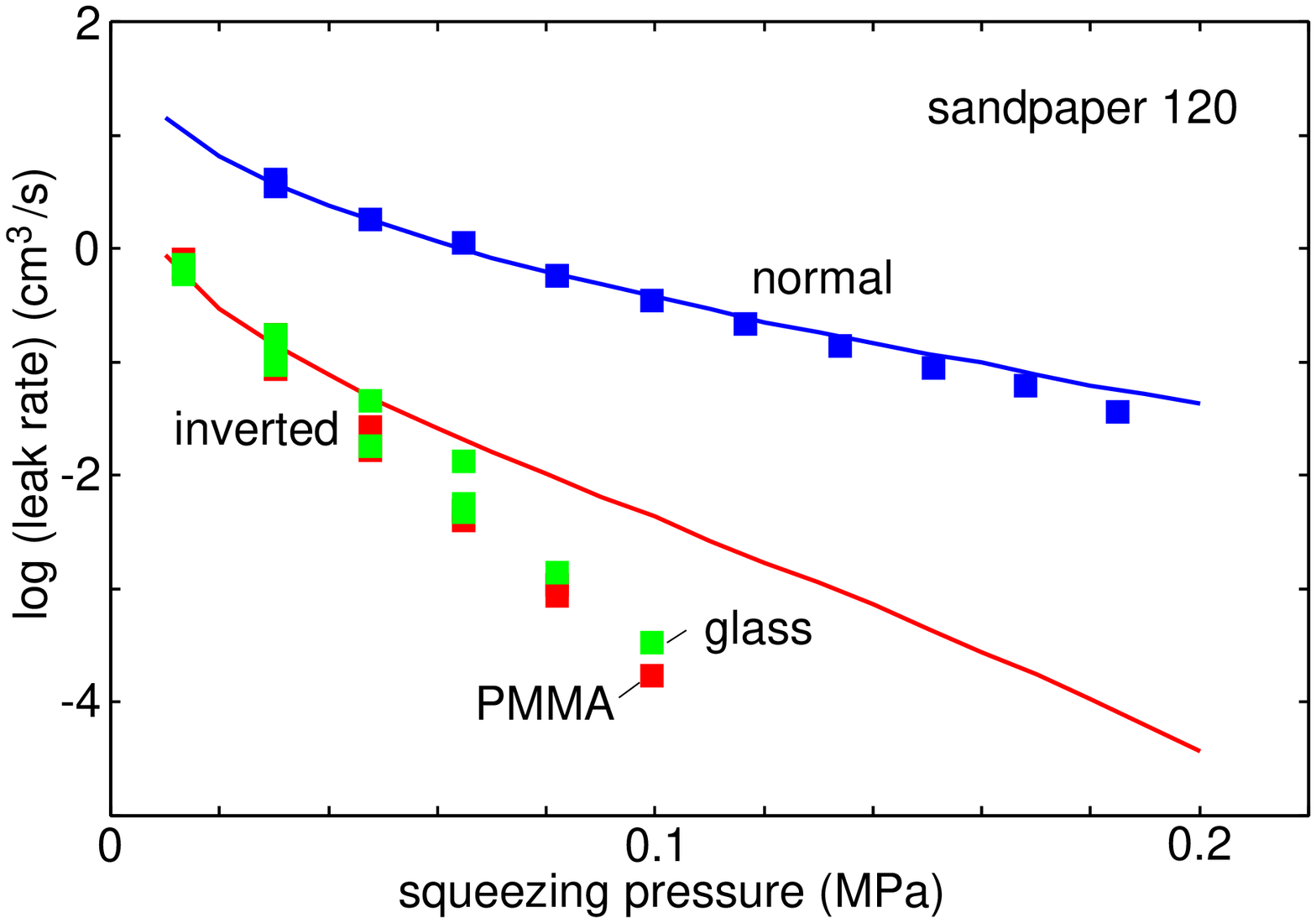}
\caption{\label{Sandpaper120.normal.alpha=1.inverted.alpha=0.8}
Square symbols: the measured leak rate for sandpaper 100 substrate (upper symbols) and for
an inverted surface (lower symbols). 
The solid lines are
the calculated leak rate using the critical-junction theory with the percolation 
threshold $p_{\rm c} = 0.6$. In the
calculation for the top curve we used the top power spectrum $C^*_{\rm T}(q)$ obtained from the
measured surface topography. For the inverted surface (bottom curve) we used the bottom
power spectrum $C^*_{\rm B}(q)$. The measured rubber elastic
modulus $E=2.3 \ {\rm MPa}$ and the fluid pressure difference 
$\Delta P = P_{\rm a}-P_{\rm b} = 10 \ {\rm kPa}$ obtained from the height of the water column.  
In the calculations we have used $\alpha = 1$ (upper curve) and $\alpha = 0.8$ (lower curve).
}
\end{figure}

In Fig. \ref{Sandpaper120.normal.alpha=1.inverted.alpha=0.8} we show similar results for
the sandpaper 100 substrate (upper symbols) and for
an inverted surface (lower symbols). Note in Fig. \ref{Sandpaper100.normal.alpha=1.inverted.alpha=0.8} 
and \ref{Sandpaper120.normal.alpha=1.inverted.alpha=0.8} the huge difference 
(roughly two orders of magnitude) between the leak-rate for the two different configurations, 
involving the original and inverted surface topographies. Note that the theory is able to describe
the observed effect if the top power spectrum is used in the analysis (which means using the bottom
power spectrum of the sandpaper surfaces in the case of the inverted surfaces). However,
for the case of the inverted surfaces the leak-rate for large enough squeezing pressure decreases
faster with the squeezing pressure than is predicted by the theory. We attribute this to the influence of adhesion
on the leak-rate. That is, the asperities of the inverted surface are quite smooth (they arise from the relative smooth
polymer (resin) film in the valleys between the particles of the original sand paper surfaces) which allow for effective
adhesion between the rubber and the glass and PMMA surfaces\cite{Manoj1}. We note here that the glass surfaces were not cleaned chemically and therefore
probably covered by nanometer thick organic contamination layers\cite{Manoj2}. 
Thus one expect a dewetting transition\cite{PVT,FBW} in the asperity contact
regions between the substrate surface and the silicon rubber surface, 
resulting in an effective adhesion which pulls the surfaces in closer contact
than expected by just the influence of the squeezing pressure. 
Preliminary calculations including adhesion indeed support this picture and will be reported on elsewhere.  

To summarize, we have compared 
experimental data with theory for the leak-rate of seals. The theory is 
based on percolation theory and a recently developed contact mechanics theory.
The experiments are for (a) silicon rubber with a smooth surface in contact with two sandpaper surfaces, 
and (b) for silicon rubber surfaces prepared by cross-linking the rubber in contact with the
sandpaper surfaces, and then squeezing the rough rubber surfaces against flat glass and PMMA surfaces.
The elastic properties of the rubber and the surface topography of the 
sandpaper and PMMA surfaces are fully characterized.
We have shown that using the top power spectrum in the theory results in good agreement between theory
and experiment.

\vskip 0.3cm

{\bf Acknowledgments}

We thank Manoj Chaudhury for useful communication about the work of adhesion (Ref. \cite{Manoj1,Manoj2}).
This work, as part of the European Science Foundation EUROCORES Program FANAS, was supported from funds 
by the DFG and the EC Sixth Framework Program, under contract N ERAS-CT-2003-980409.


\begin{thebibliography}{99}
\bibitem{Manoj1}
The work of adhesion $w=\gamma_1+\gamma_2-\gamma_{12}$ between PDMS and Plexiglas (PMMA) in water
has not been measured but can be estimated to be about $55 \ {\rm mJ/m^2}$.
\bibitem{Manoj2}
If the glass is clean, water wets it almost completely, and
work of adhesion between glass and PDMS through water may be zero or
even slightly negative, i.e. there may be a repulsion instead of
attraction. However, in general, organic contamination is strongly
adsorbed to glass: the organic molecules that float in air
and  adsorbs to surface like glass are oxidized and interact with the
surface by some strong specific interaction. They are not easy to
remove by water. In this case one may estimate the work of adhesion by 
$w\approx (1-\phi)\times w({\rm glass|water|PDMS}) +
\phi\times w({\rm hydrocarbon|water|PDMS})$, where $\phi$ is the fraction of the surface
occupied by organic contamination. Typically, one expect
$w({\rm glass|water|PDMS})\approx 0$ and 
$w({\rm hydrocarbon|water|PDMS}) \approx 80 \ {\rm mJ/m^2}$.
\end{thebibliography}
\end{document}